\documentclass[twocolumn,amssymb,fleqn]{revtex4} 
\usepackage{epsfig,amssymb,amsmath,graphicx,subfigure,hyperref  }  
\usepackage{color}

\usepackage{multirow} 
\usepackage{tabularx}

\newcommand{\be} {\begin{eqnarray}}
\newcommand{\ee} {\end{eqnarray} }

\begin{document}

\title{Nonlinear dynamics and emergent statistical regularity\\ in classical
Lennard-Jones three-body system upon disturbance}
\author{Zhenwei Yao}
\email{zyao@sjtu.edu.cn}
\affiliation{School of Physics and Astronomy, and Institute of Natural
Sciences, Shanghai Jiao Tong University, Shanghai 200240, China}
\begin{abstract}
  Understanding the deep connection of microscopic dynamics and statistical
  regularity yields insights into the foundation of statistical mechanics. In
  this work, based on the classical three-body system under the Lennard-Jones
  potential upon disturbance, we illustrated the
  elusive nonlinear dynamics in terms of the neat frequency-mixing processes,
  and revealed the emergent statistical regularity in speed distribution
  along a single particle trajectory. This
  work demonstrates the promising possibility of classical few-body models for
  exploring the fundamental questions on the interface of microscopic dynamics
  and statistical physics. 
\end{abstract}

\maketitle

\section{Introduction}

Understanding the connection of microscopic dynamics and statistical properties
is a fundamental problem related to the foundation of statistical
mechanics~\cite{li1975period,cercignani1998ludwig,Kadanoff1999,dumas2014kam,krylov2014works,neill2016ergodic}.
The fascinating development of classical dynamics since the work of Poincar\'e,
especially the discovery of deterministic chaos, shows the sensitive dependence
on initial conditions and thus the generation of practically unpredictable
behaviors~\cite{sinai1989dynamical,zwanzig2001nonequilibrium,strogatz2018nonlinear,scheck2010mechanics,yang2017classical}.
Such chaotic motion is distinct from that in Langevin equation, where the
element of randomness is introduced by adding the fluctuating force term describing
the temporally-varying impact of the environment on the object
concerned~\cite{zwanzig2001nonequilibrium}.  The seemingly random motion has been
found even in relatively simple Hamiltonian systems, including hard sphere and
rod
systems~\cite{tonks1936complete,zheng1996ergodicity,cox2000efficiently,cao2018incomplete},
and various billiard
systems~\cite{sinai1989dynamical,zaslavsky2007physics,krylov2014works}. These
findings have greatly promoted our understanding on the fundamental concept of
randomness, and have challenged the widespread viewpoint on the crucial role of
the vast number of degrees of freedom in the formation of order or regularity in the
statistical sense (statistical
regularity)~\cite{Kadanoff1999,krylov2014works,neill2016ergodic,zaslavsky2007physics}.
However, the origin of randomness in motion and the connection of random
dynamics and statistical regularity have not yet been fully
understood~\cite{Kadanoff1999,dumas2014kam,krylov2014works,neill2016ergodic}.

In this work, we explore these fundamental questions by resorting to the
classical three-body Hamiltonian system: a mechanically disturbed triplet of
particles confined on the plane under the Lennard-Jones (L-J) potential~\cite{jones1924determination}.
Specifically, we study the nonlinear dynamics from the unique perspective of
frequency spectra, and reveal the emergent statistical regularity in the
distribution of instantaneous particle speed in a single particle trajectory.
The L-J potential has been widely used to model the physical interaction in a
host of condensed matter systems~\cite{jones1924determination,yao2017topological}; its
subtle dynamical effects as revealed in several contexts have not been fully
understood~\cite{kob1995testing,xu2007excess,rutzel2003nonlinear,fukuda2017figure}.
The three-body model may be realized in experiment by various L-J particles,
such as naturally occurring noble gas atoms, coarse-grained macromolecules and
colloids~\cite{rahman1964correlations,tillack2016systematic,bienias2020exotic}.
The three-body L-J system provides an ideal platform to clarify a host of
questions, such as: How does complexity arise in the seemingly simple few-body
system? What is the connection of nonlinearity and random motion? What kind of
statistical regularity will arise in the random motion?  Elucidating the
dynamical and statistical behaviors of the three-body system may yield insights
into a variety of nonequilibrium phenomena in
crystals~\cite{halperin1978theory,strandburg1988two,mitchell2016,yao2017emergent,hwang2019direct},
especially on the collective motions in many-body
systems~\cite{vicsek2012collective,schaller2013topological,nguyen2014emergent,yao2019command}.

We combine analytical and numerical approaches to address these questions.  A
numerical approach has been proven essential to study nonlinear dynamical
problems, as pioneered by Fermi, Pasta, Ulam and Tsingou in the celebrated
numerical experiment on dozens of coupled nonlinear oscillators, which is known
as the FPUT problem ~\cite{fermi1955studies}. Recently, the nonlinear dynamics
of the harmonic three-body system has been studied using a symplectic numerical
integrator, and the featured self-driven fractional rotational diffusion has
been revealed~\cite{PhysRevLett.122.024102,PhysRevE.101.032211}. In this work,
we obtain long time trajectories of motion by high-precision numerical
integration of the equations of motion, and analyze the frequency spectra of the
kinetic energy for characterizing the nonlinear dynamics. As a characteristic of
nonlinearity, we observe the proliferation and interaction of frequencies based
on a pair of fundamental modes. Similar nonlinear phenomenon is also observed in
the linear-spring system. With the proliferation of frequencies upon stronger
disturbance, the distribution of instantaneous particle speed in a
single particle trajectory becomes fully randomized, and converges to
the regular Maxwell-Boltzmann distribution. Here, the randomization mechanism
in the few-particle system is based on the intrinsic nonlinear dynamics, which
is fundamentally different from the conventional mechanism based on frequent
collisions in systems of many
particles~\cite{maxwell1860v,cercignani1998ludwig}.


\section{Model and Method}

\begin{figure}[t!]  
\centering 
\includegraphics[width=3.45in]{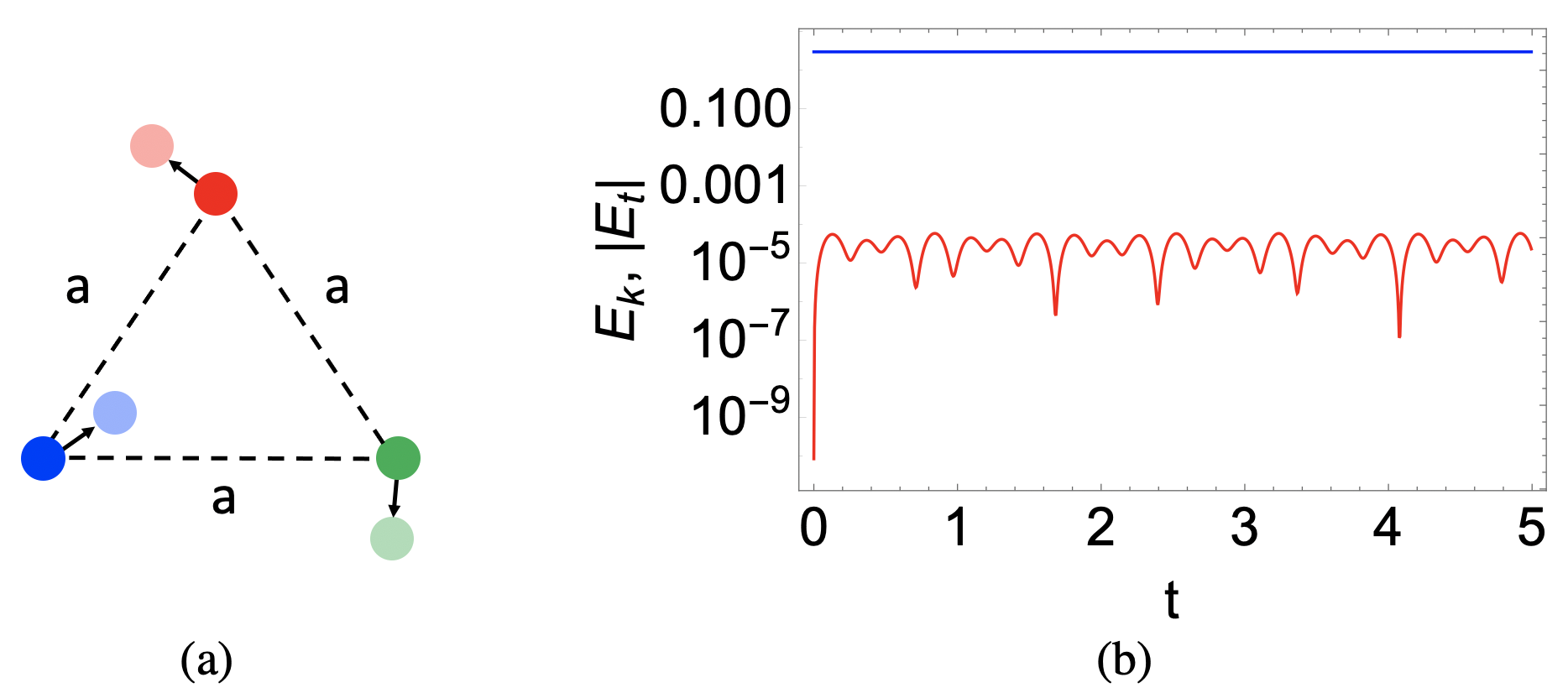}
  \caption{Typical energy variation in the perturbed three-body system under the Lennard-Jones potential. (a)
  Schematic plot of the disturbed triangular configuration of the triplet of
  particles in the plane. $a$ is the
  balance distance of the L-J potential. (b) Temporally varying kinetic energy
  (lower red curve) and the constant total energy (upper blue line). The
  disturbance amplitude $b=0.001$.  }
\label{schematic}
\end{figure}

Our model consists of a triplet of particles in the plane interacting by the L-J
potential $V_{LJ}$ [see Fig.~\ref{schematic}(a)]: 
\begin{eqnarray} 
  V_{LJ}(r) = 4\epsilon_0 \left(
(\frac{\sigma_0}{r})^{12} -  (\frac{\sigma_0}{r})^6 \right),  
\end{eqnarray} 
where $r$ is the distance between two particles, the parameters $\sigma_0$ and
$\epsilon_0$ are related to the length scale and energy scale of the L-J
potential. The potential energy has the lowest value $-\epsilon_0$ at the
balance length $a$.  $a = 2^{1/6}\sigma_0$. In this work, the units of length,
mass and time are $a$, $m$, and $\tau_0 = a \sqrt{m/\epsilon_0}$, respectively.
The initial configuration is a regular triangle of side length $a$, as shown in
Fig.~\ref{schematic}(a). The system is then disturbed.  Specifically, the
imposed displacement vector on each particle has a constant magnitude $b$ and
random orientation $\theta_i$ with respect to a reference line, where $\theta_i
\in [0, 2\pi)$ and $i=1,2,3$. For convenience in the study of the statistical
properties of the system, we also work in the frame of reference where both the
total momentum and angular momentum are zero. This could be realized by
specifying a proper initial velocity to each particle; it suffices to specify
the initial velocity $\vec{v}_{\textrm {ini}}$ to one of the three particles due
to the constraints of the conservation laws~\cite{supp2022}.  $b$ and
$v_{\textrm {ini}}$ are the key control parameters.

The microscopic dynamics of the particles upon disturbance is governed by the law of
classical mechanics. The state of the system is characterized by the
6-dimensional vectors $\vec{X}$ and $\vec{V}$. $\vec{X} = \{r_{i\mu}\}$, and
$\vec{V} = \{\dot{r}_{i\mu}\}$, where $r_{i\mu}$ is the $\mu$-component of the
displacement of particle $i$. $i=1,2,3$. $\mu= x, y$. Specifically,
$\vec{X}=(r_{1x},r_{1y},r_{2x},r_{2y},r_{3x},r_{3y})$. The sequence of the
components is the same for $\vec{V}$.  The equations of motion are cast as a
pair of first-order differential equations:
\begin{eqnarray}
  \frac{d\vec{X}(t)}{dt} &=& \vec{V}(t), \nonumber \\
  \frac{d\vec{V}(t)}{dt} &=& \vec{f}(\vec{X}). \label{eom}
\end{eqnarray}
$\vec{f}(\vec{X})$ is a 6-dimensional vector whose component $f_{i\mu}$ gives
the force on the particle $i$ in the $\mu$-direction; note that $i=1,2,3$, and
$\mu= x, y$. The sequence of the components of $\vec{f}$ is consistent with both
$\vec{X}$ and $\vec{V}$. The dependence of $\vec{f}$ on the displacement vector
$\vec{X}$ is complicated due to the highly correlated particle motions.  By
linearization at the vertices of the equilibrium configuration, the expression
for $\vec{f}(\vec{X})$ as a linear function of $\vec{X}$ could be written
formally as: 
\begin{eqnarray}
  f_{\alpha}(\vec{X}) =-\beta_0 K_{\alpha\beta} X_{\beta}, 
\end{eqnarray} 
where $\beta_0=V_{LJ}''(a)$, and the $6\times 6$ matrix $K$ contains the cofficients
of the linear terms. The matrix $K$, which is called the dynamical matrix,
governs the dynamical evolution of the perturbed system in the linear regime. To
explore the nonlinear regime, we numerically integrate Eqs.(\ref{eom}) using the
standard Verlet method at high precision and obtain accurate trajectories of
motion up to a hundred million time steps with well conserved energy, momentum
and angular momentum~\cite{rapaport2004art,supp2022}. Note that the algorithm of
Verlet integration preserves the symplectic form on phase space that guarantees
conservation of energy and momenta in long-time simulations. The typical time step
is of order of $10^{-4}\tau_0$ by striking a balance on the conservation of
relevant physical quantities and the efficiency of computation.

\section{Results and discussion}

\subsection{Dynamical analysis}

\begin{figure*}[t]  
\centering 
\includegraphics[width=6.9in]{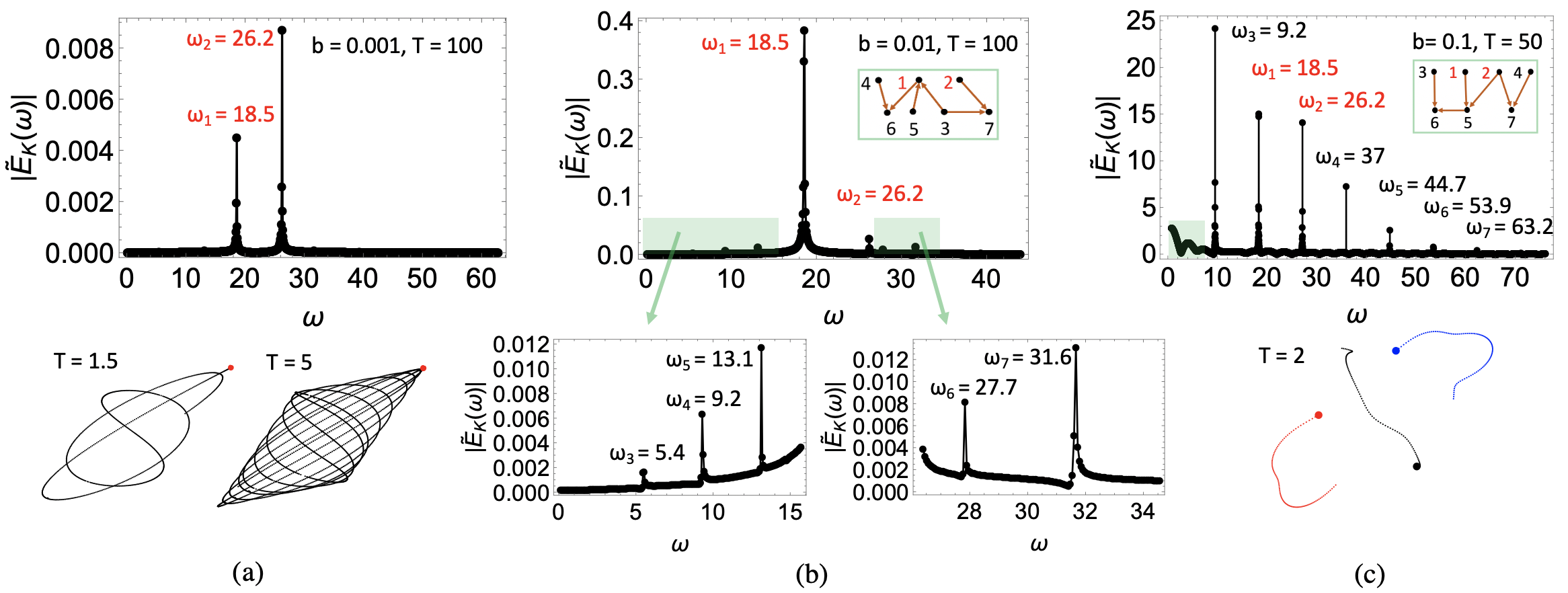}
  \caption{Frequency spectral analysis reveals nonlinear frequency-mixing
  processes in the disturbed three-body system. The frequency spectra of kinetic
  energy are obtained by discrete Fourier transformation. (a)-(c) The
  disturbance amplitude $b$ increases from $0.001$ to $0.1$.  The relation of
  the emergent frequencies $\omega_i$ is presented in the inset graphs [see the
  green boxes in (b) and (c)].  The data are recorded in the sampling interval
  of $t \in [0, T]$. Associated particle trajectories during the time intervals
  of $t\in [0, T]$ for the linear and nonlinear cases are presented in the lower
  panels in (a) and (c). 
  }
\label{vary_b}
\end{figure*}

We first slightly disturb the system, and analyze the temporal variation of the
kinetic energy in order to extract key dynamical information~\cite{yao2021epl}.
The variation of the kinetic energy in time is shown in
Fig.~\ref{schematic}(b). For reference, the constancy of the total energy is also
shown by the horizontal blue line in Fig.~\ref{schematic}(b). The associated
single particle trajectories during the time intervals of $t\in [0, 1.5]$ and
$t\in [0, 5]$ are shown in the lower panel in Fig.~\ref{vary_b}(a), where the
red dot represents the initial position of the particle. The evolving particle
trajectory weaves a highly regular parallelogrammic net bounded by four envelope
lines. Longer time trajectories are presented in SI~\cite{supp2022}.

To uncover the beat in the particle oscillations, we perform discrete Fourier
transformation on the collected energy data~\cite{sundararajan2001discrete}. A
pair of fundamental modes are revealed as shown in Fig.~\ref{vary_b}(a). $\tilde{E}_{K}(\omega)$ is
the Fourier transform of the time series of $E_k(t)$, where $t\in [0, T]$. The
values of the fundamental frequencies are $\omega_1=18.5$, and $\omega_2=26.2$. 
The Lissajous-figure-like trajectory pattern in the lower panel of
Fig.~\ref{vary_b}(a) results from a superposition of these two modes.
Simulations also show that the basic
peak structure in the $\tilde{E}_{K}(\omega)$-curve is invariant with the
variation of the orientation of the initial displacement; the peaks become
sharpened with the increase of the sampling interval $T$~\cite{supp2022}.

To explore the physical origin of the pair of the fundamental modes, we
analytically analyze the linearized equations of motion. Specifically, we
analytically derive all of the matrix elements of the dynamical matrix
$K_{\alpha\beta}$~\cite{supp2022}, and obtain the nonzero eigenvalues:
$\lambda_1 = 3/2$ and $\lambda_2=3$. Note that the first eigenvalue possesses
the two-fold degeneracy. Each eigenvalue $\lambda$ corresponds to a normal mode
of frequency $\omega_{\lambda} = \sqrt{\beta_0 \lambda/m}$. Note that these
fundamental frequencies derived in the linear regime are identical to those for
the harmonic three-body
system~\cite{PhysRevLett.122.024102,PhysRevE.101.032211}. The associated
frequency in the kinetic energy curve $E_K(t)$ is doubled:
\begin{eqnarray}
  \omega_{1} &=& 2^{\frac{11}{6}} 3^{\frac{3}{2}} \approx 18.5, \nonumber \\
  \omega_{2} &=& \sqrt{2} \omega_{1} \approx 26.2. 
\end{eqnarray} 
These analytically derived frequencies are exactly those of the
numerically solved fundamental modes in Fig.~\ref{vary_b}(a).

Here, we emphasize that the ratio of $\omega_{1}/\omega_{2}$ is an invariant for
generic confining potentials; it is independent of particle mass and potential
stiffness in equilibrium configuration. Furthermore, the two frequencies satisfy
the non-resonance condition, suggesting the existence of dynamical complexity in
the seemingly simple system~\cite{zaslavsky2007physics}. To substantiate this
point, we record the crossing points of the evolving trajectory in
Fig.~\ref{vary_b}(a) with some reference line up to $T=50$ following the spirit
of Poincar\'e map, and numerically observe the filling of space in time by the aperiodic
trajectory~\cite{supp2022,Kadanoff1999,altmann2006stickiness,scheck2010mechanics,krylov2014works}.

Upon stronger disturbance, richer frequency spectra are observed as presented in
Figs.~\ref{vary_b}(b) and \ref{vary_b}(c) for $b=0.01$ and $b=0.1$,
respectively. As a characteristic of nonlinearity, we observe the proliferation
of frequencies on the basis of the two fundamental frequencies $\omega_1$ and
$\omega_2$~\cite{fermi1955studies}. By tracking the temporal variation of the
peaks, we find that the dynamical state of the system is initially dominated by
the single $\omega_1$ mode. The ensuing growth of the height of the peaks
reflects the complicated partition of energy among the modes~\cite{supp2022}.

The connection of each emergent frequency and the two fundamental frequencies
($\omega_1$ and $\omega_2$) is illustrated in the inset graphs in
Figs.~\ref{vary_b}(b) and \ref{vary_b}(c). A vertex $i$ represents frequency
$\omega_i$. The frequency at a vertex with a pair of incoming arrows is either
the sum or the difference of the two frequencies at the ends of the arrows.
Remarkably, a simple classical mechanical system consisting of only three
particles is capable of supporting a series of complicated frequency-mixing
processes, including the sum-frequency and difference-frequency generation. In
Fig.~\ref{vary_b}(b), we further observe the fission
of the fundamental mode $\omega_1$ into a pair of lower-frequency modes
($\omega_1\rightarrow \omega_3$ and $\omega_5$), and the frequency-halving
process ($\omega_1 \rightarrow \omega_4$, and $\omega_2\rightarrow \omega_5$)
that is related to the bifurcation of orbits~\cite{scheck2010mechanics}. Similar frequency-halving
phenomena have been reported in periodically poled active nonlinear
crystals~\cite{chirkin2004nonclassical}.

Comparison of Figs.~\ref{vary_b}(b) and \ref{vary_b}(c) shows the enhanced
nonlinear effect as the disturbance amplitude $b$ is increased. Increasing $b$
leads to significant growth of the peak height and emergence of higher-frequency
modes. Notably, secondary frequency-mixing events are observed in
Fig.~\ref{vary_b}(c). For example, the combination of $\omega_3$ and $\omega_5$
leads to the new high-frequency $\omega_6$ mode. We also notice the accumulation
of low-frequency modes to form a continuous frequency spectrum as indicated by
the light green box in Fig.~\ref{vary_b}(c). Furthermore, the seemingly free
motions of the particles, as plotted in the lower panel in Fig.~\ref{vary_b}(c),
are highly correlated by the conservation of energy, momentum and angular
momentum.

In preceding discussions, the particles interact by the L-J potential, which
could be regarded as a nonlinear spring. To examine the generality of
the key observations about the neat frequency-mixing events, we further study
the triplet system connected by linear springs~\cite{supp2022}. In the
perturbation regime, we observe a pair of fundamental modes that are identical
to that in the L-J system as expected. At $b=0.1$ [corresponding to the case in
Fig.~\ref{vary_b}(c)], we also observe the proliferation of new frequencies (but
with much weaker strength as compared with the L-J system) and the featured
frequency-mixing processes in the linear-spring system. The frequency spectra
and relevant trajectories of the linear-spring system are presented in Supplemental
Materials~\cite{supp2022}. These results indicate the
existence of nonlinear dynamics even for a harmonic potential, implying the
richness of the three-body system in dynamics.

It is known that the generation
of new frequencies could be attributed to the nonlinear effect, which can be
understood by analyzing the nonlinear equation of motion~\cite{Landau_mechanics}
or by resorting to logistic mapping~\cite{Feigenbaum1978}. In the specific
three-body system under harmonic interaction, where the nonlinear effect as
characterized by the proliferation of new frequencies also arises, the nonlinear
interaction originates from the geometric configuration of the three particles,
which is known as the geometric nonlinearity~\cite{de2009relating,PhysRevLett.122.024102}.
The presence of a third particle brings in the nonlinear element that changes
the original frequency structure of the original two-body system. 
Simulations show that the nonlinear effect vanishes by aligning the particles
along a straight line. Thus, it shall be emphasized that the nonlinearity
requires a non-zero balance length and placing the particles in a plane and not
on a line.  Here, we shall mention that the nonlinear dynamics of the harmonic
three-body system has been discussed from the perspective of rotational random
walk, where the spectral structure of the motion in terms of particle position
(angular displacement or x coordinate) has been discussed, and the featured
behavior of self-driven fractional rotational diffusion has been
revealed~\cite{PhysRevLett.122.024102,PhysRevE.101.032211}.

Regarding the free motions of the particles as constrained by the conservation
laws in Fig.~\ref{vary_b}(c), it is of interest to mention that the conservation
laws admit the counter-intuitive phenomenon of ejecting a particle to infinity
in finite time in the Newtonian N-body system, as rigorously proved in
mathematics~\cite{saari1995off}. This phenomenon has motivated several deep
mathematical conclusions related to the issue of singularity structure in
classical mechanical systems.

\begin{figure*}[th]  
\centering 
\includegraphics[width=7in]{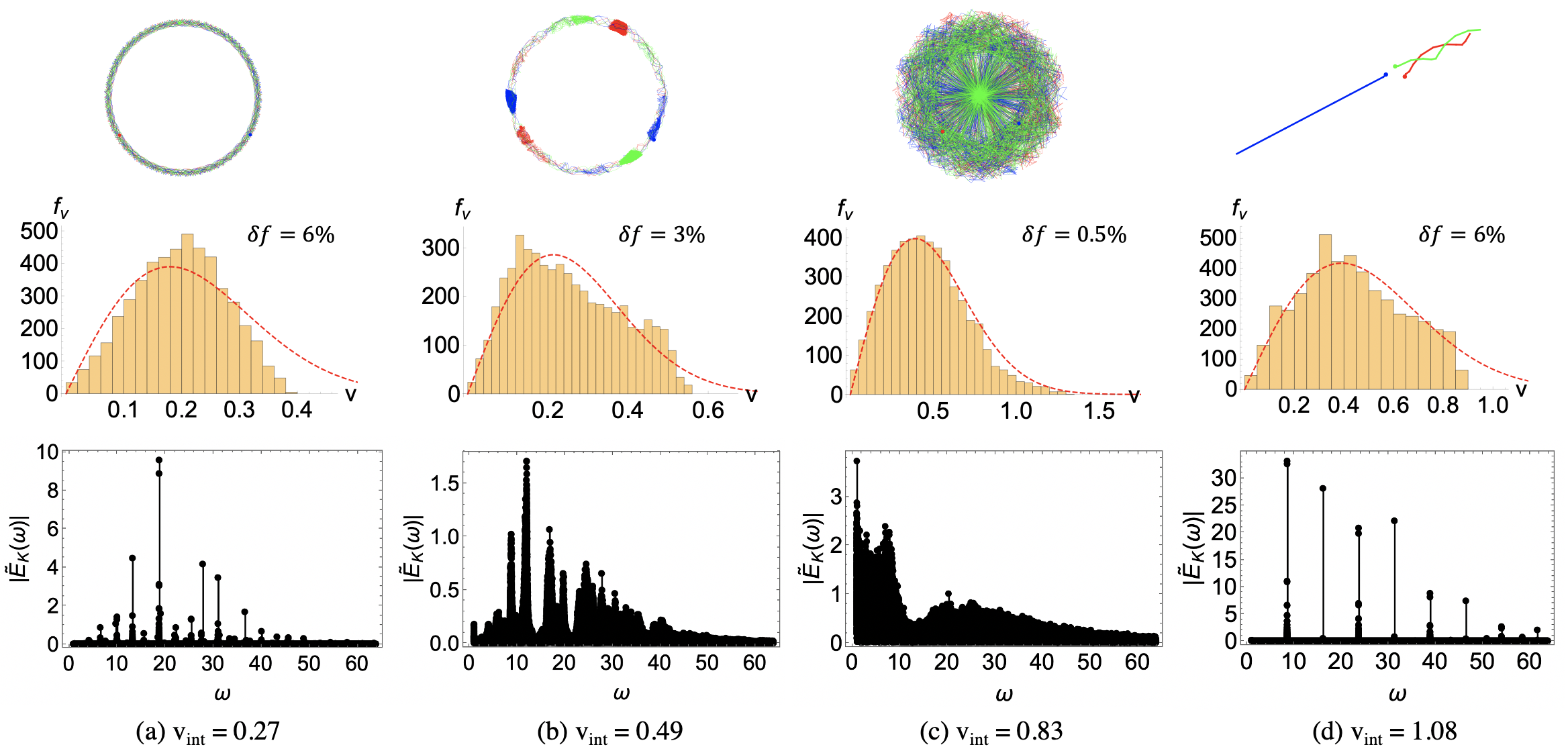}
  \caption{Statistical regularity arises in the classical three-body system
  under enhanced nonlinear effect. The histograms show the distribution of the
  instantaneous speed along a single particle trajectory at varying initial
  speed $v_{\textrm {ini}}$. For bounded motions in (a)-(c), the speed
  distribution approaches the Maxwell-Boltzmann distribution as $v_{\textrm
  {ini}}$ is increased. Note that this trend applies to all of the three
  particles. The fitting functions in the form of the
  Maxwell-Boltzmann distribution are plotted in dashed red curves.  $\delta f$
  is a measure of the deviation of the numerically obtained speed distribution
  from the Maxwell-Boltzmann distribution. Further increasing $v_{\textrm
  {ini}}$ leads to the free motion of the particles as shown in (d). The
  associated particle trajectories and the frequency spectra of kinetic energy
  are also presented; the three particles are distinguished by different colors.
  The particle speed is recorded at equal time interval $\Delta t = 5$ up to
  $T=25000$.    
}
\label{mb}
\end{figure*}

\begin{figure}[t!]  
\centering 
\includegraphics[width=3.7in]{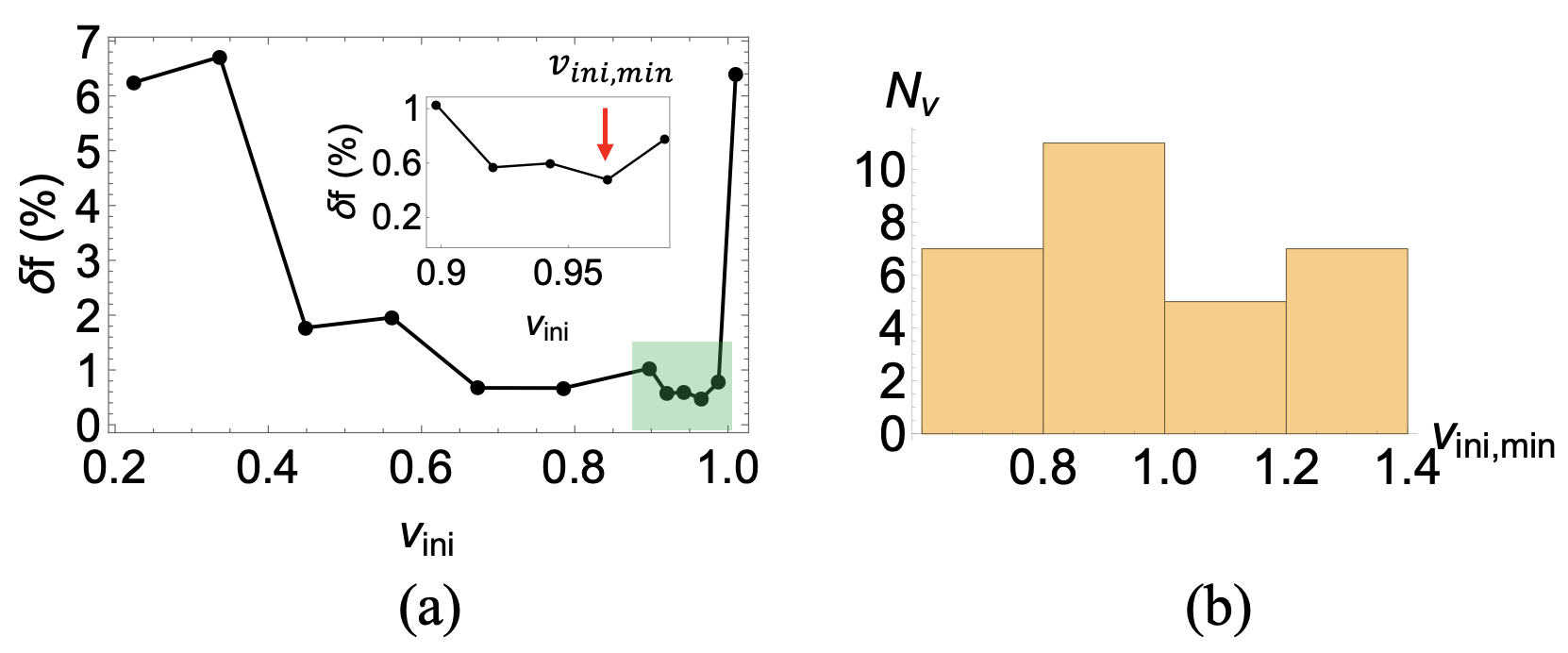}
  \caption{Statistical analysis on the convergence of the speed distribution to
  the two-dimensional Maxwell-Boltzmann distribution in the bounded motion under
  varying initial conditions. (a) A typical plot of $\delta f$ versus
  $v_{ini}$. $\delta f$ is the deviation between the numerically obtained speed
  distribution and the Maxwell-Boltzmann distribution. $v_{ini}$ is the
  magnitude of initial speed.  The region in the green highlight is enlarged in
  the inset. $\delta f$ reaches minimum $\delta f_{min}$ when
  $v_{ini}=v_{ini,min}$. Dynamical transition from bounded to unbounded state
  occurs at the right end of the plot as $v_{ini}$ exceeds about $0.98$. (b)
  Histogram of $v_{ini,min}$ based on the data of particle speed under 30
  independent initial conditions.  $v_{ini,min}=0.97 \pm 0.22$. The mean value
  of $\delta f_{min}$ is as small as $0.57\%$, indicating the good agreement
  between the numerical results and the Maxwell-Boltzmann distribution.}
\label{mb_reply}
\end{figure}

\subsection{Statistical analysis}
In the nonlinear regime, with the reduction of the regularity of motion due to
the proliferation of frequencies, could any statistical regularity arise? In
classical ideal gas, via frequent molecular collisions, the speed of the gas
particles approaches the Maxwell-Boltzmann distribution as governed by the
central limit theorem in the thermodynamic
limit~\cite{maxwell1860v,krylov2014works}. Note that non-ergodic, stable motions
of $N$ repulsive particles confined to a compact $d$-dimensional ($d\leq 2$)
domain at sufficiently large energy is reported recently~\cite{rom2022stable}.
Here, in the classical few-body system, could the nonlinearity effect fully
randomize the particle speed to realize the distribution as in equilibrium
systems of many particles?

To address this question, we work in the frame of reference where both the total
momentum and angular momentum are zero. The initial condition is formulated by
specifying the initial velocities of the three particles~\cite{supp2022}. Our
statistical analysis is based on the single particle trajectory without
resorting to the procedures of ensemble averaging and coarse-graining. We focus
on the distribution of speed along a trajectory. In other words, the randomness
of motion is discussed in the sense of speed distribution. The distributions of
the speed data recorded at equal time interval are presented in Fig.~\ref{mb}.
The associated trajectories of motion and the frequency spectra of kinetic
energy are also shown in Fig.~\ref{mb}.

In the bounded motion of particles from Fig.~\ref{mb}(a) to \ref{mb}(c), it is
found that the speed distribution converges to the two-dimensional
Maxwell-Boltzmann distribution, as indicated by the dashed red curve in
Fig.~\ref{mb}(c).  This trend applies to all of the three particles.  To measure
the deviation between the numerically obtained speed distribution and the
Maxwell-Boltzmann distribution, we propose the quantity $\delta f$ as defined
below:
\begin{eqnarray}
  \delta f = \frac{\sum_{v_i} (f(v_i) - f_{M-B}(v_i))^2}{\sum_{v_i} f_{M-B}(v_i)^2}. 
\end{eqnarray}
$f(v)$ is the numerically obtained speed distribution, and $f_{M-B}(v)$ is the
Maxwell-Boltzmann distribution.
\begin{eqnarray}
  f_{M-B}(v) \delta v = \Gamma v
\exp(-v^2/2v_p^2) \delta v, 
\end{eqnarray}
where $v_p$ is the most probable speed. The values
of $\delta f$ are given alongside the histograms in Fig.~\ref{mb}. The
asymptotic process towards the speed distribution in Fig.~\ref{mb}(c) with the
increase of the sampling interval $T$ is presented in SI~\cite{supp2022}. The
value of $\delta f$ monotonously decreases from $20\%$ to $0.7\%$ when $T$
increases from $500$ to $10000$.

Simulations show that under varying initial conditions, the speed distribution
uniformly converges to the two-dimensional Maxwell-Boltzmann distribution in the
bounded motion with the increase of $v_{ini}$. Figure~\ref{mb_reply}(a) shows a
typical plot of $\delta f$ versus $v_{ini}$. We see that with the increase of
$v_{ini}$, the value of $\delta f$ tends to decrease in the regime of bounded
motion; dynamical transition from bounded to unbounded state occurs at the right
end of the plot as $v_{ini}$ exceeds about $0.98$. The value of $\delta f$
reaches minimum when $v_{ini}=v_{ini,min}$, as indicated in the inset for the
zoomed-in region in green highlight. For a collection of random orientations of
the initial velocity, it is found that the value of $v_{ini}=v_{ini,min}$ is
subject to a fluctuation: $v_{ini,min}=0.97 \pm 0.22$, based on the data of
particle speed under 30 independent initial conditions. The histogram of
$v_{ini,min}$ is presented in Fig.~\ref{mb_reply}(b). The mean value of $\delta
f_{min}$ is as small as $0.57\%$, indicating the good agreement between the
numerical results and the Maxwell-Boltzmann distribution.

Further increasing $v_{\textrm{ini}}$ leads to the dynamical transition from
bounded to unbounded state. In the bounded motion, the region of motion evolves
from an annulus [Fig.~\ref{mb}(a)] to a filled circle [Fig.~\ref{mb}(c)].
Notably, the trajectory in Fig.~\ref{mb}(b) shows simultaneous long-duration
trapping of the triplet of particles within small angular intervals;
quantitative analysis of this phenomenon is presented in SI~\cite{supp2022}.
Such a L$\acute{e}$vy-flight-like phenomenon is related to the stickiness
phenomenon of chaotic trajectories in the phase space of a Hamiltonian
system~\cite{altmann2006stickiness,zaslavsky2007physics}, and it has also been
observed in the harmonic three-mass
system~\cite{PhysRevLett.122.024102,PhysRevE.101.032211}.  The unbounded state
in Fig.~\ref{mb}(d) is featured with the intertwined motion of the pair of
particles and the ballistic motion of the remaining one. The associated
frequency spectrum shows that such a dynamical state is dominated by a few
discrete frequencies.

Comparison of the four cases in Fig.~\ref{mb} shows that the Maxwell-Boltzmann
distribution is reached when the trajectories of the three particles are fully
mixed (see the upper insets), or when sufficiently large number of new
frequencies are generated from the perspective of frequency spectra (see the
lower insets). In other words, a great number of collisions among the particles
are crucial for shaping the Maxwell-Boltzmann distribution along a single
particle trajectory. Note that in the collisions of the particles, the large
repulsive force between particles in short distance requires a sufficiently fine
time step to reduce the particle displacement in a time step and to ensure the
conservation of energy. Simulations show that the system energy is well
conserved for the adopted time step at the order of $10^{-4}\tau_0$~\cite{supp2022}. For
the case that the speed of one particle is much larger than the remaining ones,
as shown in Fig.~\ref{mb}(d), the fast particle may escape.  Consequently, the
trajectories could not be fully mixed, and the speed distribution is deviated
from the Maxwell-Boltzmann distribution  at least during the given simulation
time.  Note that even in this unbounded dynamical state, the nonlinear effect as
characterized by the proliferation of frequencies is still present [see the
lower panel in Fig.~\ref{mb}(d) for the spectral structure].

It shall be emphasized that the revealed statistical regularity in the speed
distribution is neither resorting to any ensemble averaging nor by
coarse-graining procedure. The element of randomness, which is crucial for
shaping the regular statistical law in the few-body system, originates from
the intrinsic dynamical nonlinearity without any external stimulus. Here,
frequency spectrum analysis serves as a powerful tool to extract important
nonlinear physics. Specifically, the nonlinearity caused frequency-mixing
processes fully randomize the motion of the particles, and generate the
regular distribution of instantaneous particle speed. We also note that the
randomness of the motion is in the sense of the distribution of speed along a
single trajectory, but not in the shape of the trajectory. This randomization
mechanism is fundamentally different from the relaxation mechanism based on
frequent molecular collisions in many-body
systems~\cite{maxwell1860v,cercignani1998ludwig}. This finding may advance our
understanding on the connection of classical dynamics and statistical mechanics.
The foundation of statistical mechanics is usually assumed to be based on the
presence of a large number of constituents and on
ergodicity~\cite{Kadanoff1999,scheck2010mechanics,krylov2014works}, which has
been challenged recently~\cite{epjst2015}. The present system consists of only
three particles, and its dynamics is governed by the deterministic law. Yet,
statistical regularity arises. Namely, the speed distribution along a single
particle trajectory displays a good agreement with the regular Maxwell-Boltzmann
distribution.

\section{Conclusion}

In summary, this work demonstrates how repeated application of simple laws leads
to spatio-temporal complexity in a classical mechanical system consisting of
only three particles. We have illustrated the elusive nonlinear dynamics and the
connection of nonlinearity and random motion in terms of the neat
frequency-mixing processes. We have also revealed the statistical regularity in
particle speed underlying a single particle trajectory. The elucidation of the
nonlinear dynamics and the emergent statistical regularity in this work may
inspire further explorations into the fundamental questions on the interface of
microscopic dynamics and statistical physics. Especially, the formally simple
three-body system may serve as a suitable model to address these
questions~\cite{PhysRevLett.122.024102,PhysRevE.101.032211}.  A possible extension of the current
work is to endow the particles with distinct masses ($m_i$ for particle $i$),
which brings in multiple time scales ($\tau_i=a\sqrt{m_i/\epsilon_0}$) and thus
a wealth of dynamical behaviors arising from the interplay of these new time
scales.  \\

\section{ACKNOWLEDGMENTS}

This work was supported by the National Natural Science Foundation of China
(Grants No. BC4190050).


\begin{thebibliography}{50}
\expandafter\ifx\csname natexlab\endcsname\relax\def\natexlab#1{#1}\fi
\expandafter\ifx\csname bibnamefont\endcsname\relax
  \def\bibnamefont#1{#1}\fi
\expandafter\ifx\csname bibfnamefont\endcsname\relax
  \def\bibfnamefont#1{#1}\fi
\expandafter\ifx\csname citenamefont\endcsname\relax
  \def\citenamefont#1{#1}\fi
\expandafter\ifx\csname url\endcsname\relax
  \def\url#1{\texttt{#1}}\fi
\expandafter\ifx\csname urlprefix\endcsname\relax\def\urlprefix{URL }\fi
\providecommand{\bibinfo}[2]{#2}
\providecommand{\eprint}[2][]{\url{#2}}

\bibitem[{\citenamefont{Li and Yorke}(1975)}]{li1975period}
\bibinfo{author}{\bibfnamefont{T.-Y.} \bibnamefont{Li}} \bibnamefont{and}
  \bibinfo{author}{\bibfnamefont{J.~A.} \bibnamefont{Yorke}},
  \bibinfo{journal}{Am. Math. Mon.} \textbf{\bibinfo{volume}{82}},
  \bibinfo{pages}{985} (\bibinfo{year}{1975}).

\bibitem[{\citenamefont{Cercignani et~al.}(1998)}]{cercignani1998ludwig}
\bibinfo{author}{\bibfnamefont{C.}~\bibnamefont{Cercignani}}
  \bibnamefont{et~al.}, \emph{\bibinfo{title}{Ludwig Boltzmann: The Man Who
  Trusted Atoms}} (\bibinfo{publisher}{Oxford University Press, Oxford},
  \bibinfo{year}{1998}).

\bibitem[{\citenamefont{Kadanoff}(1999)}]{Kadanoff1999}
\bibinfo{author}{\bibfnamefont{L.~P.} \bibnamefont{Kadanoff}},
  \emph{\bibinfo{title}{From Order to Chaos II}} (\bibinfo{publisher}{World
  Scientific}, \bibinfo{year}{1999}).

\bibitem[{\citenamefont{Dumas}(2014)}]{dumas2014kam}
\bibinfo{author}{\bibfnamefont{H.~S.} \bibnamefont{Dumas}},
  \emph{\bibinfo{title}{The KAM Story}} (\bibinfo{publisher}{World Scientific
  Publishing Company}, \bibinfo{year}{2014}).

\bibitem[{\citenamefont{Krylov}(2014)}]{krylov2014works}
\bibinfo{author}{\bibfnamefont{N.~S.} \bibnamefont{Krylov}},
  \emph{\bibinfo{title}{Works On The Foundations of Statistical Physics}}
  (\bibinfo{publisher}{Princeton University Press}, \bibinfo{year}{2014}).

\bibitem[{\citenamefont{Neill et~al.}(2016)\citenamefont{Neill, Roushan, Fang,
  Chen, Kolodrubetz, Chen, Megrant, Barends, Campbell, Chiaro
  et~al.}}]{neill2016ergodic}
\bibinfo{author}{\bibfnamefont{C.}~\bibnamefont{Neill}},
  \bibinfo{author}{\bibfnamefont{P.}~\bibnamefont{Roushan}},
  \bibinfo{author}{\bibfnamefont{M.}~\bibnamefont{Fang}},
  \bibinfo{author}{\bibfnamefont{Y.}~\bibnamefont{Chen}},
  \bibinfo{author}{\bibfnamefont{M.}~\bibnamefont{Kolodrubetz}},
  \bibinfo{author}{\bibfnamefont{Z.}~\bibnamefont{Chen}},
  \bibinfo{author}{\bibfnamefont{A.}~\bibnamefont{Megrant}},
  \bibinfo{author}{\bibfnamefont{R.}~\bibnamefont{Barends}},
  \bibinfo{author}{\bibfnamefont{B.}~\bibnamefont{Campbell}},
  \bibinfo{author}{\bibfnamefont{B.}~\bibnamefont{Chiaro}},
  \bibnamefont{et~al.}, \bibinfo{journal}{Nat. Phys.}
  \textbf{\bibinfo{volume}{12}}, \bibinfo{pages}{1037} (\bibinfo{year}{2016}).

\bibitem[{\citenamefont{Sinai}(1989)}]{sinai1989dynamical}
\bibinfo{author}{\bibfnamefont{I.~G.} \bibnamefont{Sinai}},
  \emph{\bibinfo{title}{Dynamical Systems II}} (\bibinfo{publisher}{Springer},
  \bibinfo{year}{1989}).

\bibitem[{\citenamefont{Zwanzig}(2001)}]{zwanzig2001nonequilibrium}
\bibinfo{author}{\bibfnamefont{R.}~\bibnamefont{Zwanzig}},
  \emph{\bibinfo{title}{Nonequilibrium Statistical Mechanics}}
  (\bibinfo{publisher}{Oxford University Press, USA}, \bibinfo{year}{2001}).

\bibitem[{\citenamefont{Strogatz}(2018)}]{strogatz2018nonlinear}
\bibinfo{author}{\bibfnamefont{S.~H.} \bibnamefont{Strogatz}},
  \emph{\bibinfo{title}{Nonlinear Dynamics and Chaos}} (\bibinfo{publisher}{CRC
  Press}, \bibinfo{year}{2018}).

\bibitem[{\citenamefont{Scheck}(2010)}]{scheck2010mechanics}
\bibinfo{author}{\bibfnamefont{F.}~\bibnamefont{Scheck}},
  \emph{\bibinfo{title}{Mechanics: from Newton's Laws to Deterministic Chaos}}
  (\bibinfo{publisher}{Springer Science \& Business Media},
  \bibinfo{year}{2010}).

\bibitem[{\citenamefont{Yang et~al.}(2017)\citenamefont{Yang,
  P{\'e}rez-R{\'\i}os, and Robicheaux}}]{yang2017classical}
\bibinfo{author}{\bibfnamefont{B.}~\bibnamefont{Yang}},
  \bibinfo{author}{\bibfnamefont{J.}~\bibnamefont{P{\'e}rez-R{\'\i}os}},
  \bibnamefont{and}
  \bibinfo{author}{\bibfnamefont{F.}~\bibnamefont{Robicheaux}},
  \bibinfo{journal}{Phys. Rev. Lett.} \textbf{\bibinfo{volume}{118}},
  \bibinfo{pages}{154101} (\bibinfo{year}{2017}).

\bibitem[{\citenamefont{Tonks}(1936)}]{tonks1936complete}
\bibinfo{author}{\bibfnamefont{L.}~\bibnamefont{Tonks}},
  \bibinfo{journal}{Phys. Rev.} \textbf{\bibinfo{volume}{50}},
  \bibinfo{pages}{955} (\bibinfo{year}{1936}).

\bibitem[{\citenamefont{Zheng et~al.}(1996)\citenamefont{Zheng, Hu, and
  Zhang}}]{zheng1996ergodicity}
\bibinfo{author}{\bibfnamefont{Z.}~\bibnamefont{Zheng}},
  \bibinfo{author}{\bibfnamefont{G.}~\bibnamefont{Hu}}, \bibnamefont{and}
  \bibinfo{author}{\bibfnamefont{J.}~\bibnamefont{Zhang}},
  \bibinfo{journal}{Phys. Rev. E} \textbf{\bibinfo{volume}{53}},
  \bibinfo{pages}{3246} (\bibinfo{year}{1996}).

\bibitem[{\citenamefont{Cox and Ackland}(2000)}]{cox2000efficiently}
\bibinfo{author}{\bibfnamefont{S.}~\bibnamefont{Cox}} \bibnamefont{and}
  \bibinfo{author}{\bibfnamefont{G.}~\bibnamefont{Ackland}},
  \bibinfo{journal}{Phys. Rev. Lett.} \textbf{\bibinfo{volume}{84}},
  \bibinfo{pages}{2362} (\bibinfo{year}{2000}).

\bibitem[{\citenamefont{Cao et~al.}(2018)\citenamefont{Cao, Bulchandani, and
  Moore}}]{cao2018incomplete}
\bibinfo{author}{\bibfnamefont{X.}~\bibnamefont{Cao}},
  \bibinfo{author}{\bibfnamefont{V.~B.} \bibnamefont{Bulchandani}},
  \bibnamefont{and} \bibinfo{author}{\bibfnamefont{J.~E.} \bibnamefont{Moore}},
  \bibinfo{journal}{Phys. Rev. Lett.} \textbf{\bibinfo{volume}{120}},
  \bibinfo{pages}{164101} (\bibinfo{year}{2018}).

\bibitem[{\citenamefont{Zaslavsky}(2007)}]{zaslavsky2007physics}
\bibinfo{author}{\bibfnamefont{G.~M.} \bibnamefont{Zaslavsky}},
  \emph{\bibinfo{title}{The Physics of Chaos in Hamiltonian Systems}}
  (\bibinfo{publisher}{World Scientific}, \bibinfo{year}{2007}).

\bibitem[{\citenamefont{Jones}(1924)}]{jones1924determination}
\bibinfo{author}{\bibfnamefont{J.~E.} \bibnamefont{Jones}},
  \bibinfo{journal}{Proc. R. Soc. London, Ser. A.}
  \textbf{\bibinfo{volume}{106}}, \bibinfo{pages}{463} (\bibinfo{year}{1924}).

\bibitem[{\citenamefont{Yao}(2017{\natexlab{a}})}]{yao2017topological}
\bibinfo{author}{\bibfnamefont{Z.}~\bibnamefont{Yao}}, \bibinfo{journal}{Soft
  Matter} \textbf{\bibinfo{volume}{13}}, \bibinfo{pages}{5905}
  (\bibinfo{year}{2017}{\natexlab{a}}).

\bibitem[{\citenamefont{Kob and Andersen}(1995)}]{kob1995testing}
\bibinfo{author}{\bibfnamefont{W.}~\bibnamefont{Kob}} \bibnamefont{and}
  \bibinfo{author}{\bibfnamefont{H.~C.} \bibnamefont{Andersen}},
  \bibinfo{journal}{Phys. Rev. E} \textbf{\bibinfo{volume}{51}},
  \bibinfo{pages}{4626} (\bibinfo{year}{1995}).

\bibitem[{\citenamefont{Xu et~al.}(2007)\citenamefont{Xu, Wyart, Liu, and
  Nagel}}]{xu2007excess}
\bibinfo{author}{\bibfnamefont{N.}~\bibnamefont{Xu}},
  \bibinfo{author}{\bibfnamefont{M.}~\bibnamefont{Wyart}},
  \bibinfo{author}{\bibfnamefont{A.~J.} \bibnamefont{Liu}}, \bibnamefont{and}
  \bibinfo{author}{\bibfnamefont{S.~R.} \bibnamefont{Nagel}},
  \bibinfo{journal}{Phys. Rev. Lett.} \textbf{\bibinfo{volume}{98}},
  \bibinfo{pages}{175502} (\bibinfo{year}{2007}).

\bibitem[{\citenamefont{R{\"u}tzel et~al.}(2003)\citenamefont{R{\"u}tzel, Lee,
  and Raman}}]{rutzel2003nonlinear}
\bibinfo{author}{\bibfnamefont{S.}~\bibnamefont{R{\"u}tzel}},
  \bibinfo{author}{\bibfnamefont{S.~I.} \bibnamefont{Lee}}, \bibnamefont{and}
  \bibinfo{author}{\bibfnamefont{A.}~\bibnamefont{Raman}},
  \bibinfo{journal}{Proc. R. Soc. London, Ser. A}
  \textbf{\bibinfo{volume}{459}}, \bibinfo{pages}{1925} (\bibinfo{year}{2003}).

\bibitem[{\citenamefont{Fukuda et~al.}(2017)\citenamefont{Fukuda, Fujiwara, and
  Ozaki}}]{fukuda2017figure}
\bibinfo{author}{\bibfnamefont{H.}~\bibnamefont{Fukuda}},
  \bibinfo{author}{\bibfnamefont{T.}~\bibnamefont{Fujiwara}}, \bibnamefont{and}
  \bibinfo{author}{\bibfnamefont{H.}~\bibnamefont{Ozaki}}, \bibinfo{journal}{J.
  Phys. A: Math. Theor.} \textbf{\bibinfo{volume}{50}}, \bibinfo{pages}{105202}
  (\bibinfo{year}{2017}).

\bibitem[{\citenamefont{Rahman}(1964)}]{rahman1964correlations}
\bibinfo{author}{\bibfnamefont{A.}~\bibnamefont{Rahman}},
  \bibinfo{journal}{Phys. Rev.} \textbf{\bibinfo{volume}{136}},
  \bibinfo{pages}{A405} (\bibinfo{year}{1964}).

\bibitem[{\citenamefont{Tillack et~al.}(2016)\citenamefont{Tillack, Johnson,
  Eichinger, and Robinson}}]{tillack2016systematic}
\bibinfo{author}{\bibfnamefont{A.~F.} \bibnamefont{Tillack}},
  \bibinfo{author}{\bibfnamefont{L.~E.} \bibnamefont{Johnson}},
  \bibinfo{author}{\bibfnamefont{B.~E.} \bibnamefont{Eichinger}},
  \bibnamefont{and} \bibinfo{author}{\bibfnamefont{B.~H.}
  \bibnamefont{Robinson}}, \bibinfo{journal}{J. Chem. Theory Comput.}
  \textbf{\bibinfo{volume}{12}}, \bibinfo{pages}{4362} (\bibinfo{year}{2016}).

\bibitem[{\citenamefont{Bienias et~al.}(2020)\citenamefont{Bienias, Gullans,
  Kalinowski, Craddock, Ornelas-Huerta, Rolston, Porto, and
  Gorshkov}}]{bienias2020exotic}
\bibinfo{author}{\bibfnamefont{P.}~\bibnamefont{Bienias}},
  \bibinfo{author}{\bibfnamefont{M.~J.} \bibnamefont{Gullans}},
  \bibinfo{author}{\bibfnamefont{M.}~\bibnamefont{Kalinowski}},
  \bibinfo{author}{\bibfnamefont{A.~N.} \bibnamefont{Craddock}},
  \bibinfo{author}{\bibfnamefont{D.~P.} \bibnamefont{Ornelas-Huerta}},
  \bibinfo{author}{\bibfnamefont{S.~L.} \bibnamefont{Rolston}},
  \bibinfo{author}{\bibfnamefont{J.}~\bibnamefont{Porto}}, \bibnamefont{and}
  \bibinfo{author}{\bibfnamefont{A.~V.} \bibnamefont{Gorshkov}},
  \bibinfo{journal}{Phys. Rev. Lett.} \textbf{\bibinfo{volume}{125}},
  \bibinfo{pages}{093601} (\bibinfo{year}{2020}).

\bibitem[{\citenamefont{Halperin and Nelson}(1978)}]{halperin1978theory}
\bibinfo{author}{\bibfnamefont{B.}~\bibnamefont{Halperin}} \bibnamefont{and}
  \bibinfo{author}{\bibfnamefont{D.~R.} \bibnamefont{Nelson}},
  \bibinfo{journal}{Phys. Rev. Lett.} \textbf{\bibinfo{volume}{41}},
  \bibinfo{pages}{121} (\bibinfo{year}{1978}).

\bibitem[{\citenamefont{Strandburg}(1988)}]{strandburg1988two}
\bibinfo{author}{\bibfnamefont{K.~J.} \bibnamefont{Strandburg}},
  \bibinfo{journal}{Rev. Mod. Phys.} \textbf{\bibinfo{volume}{60}},
  \bibinfo{pages}{161} (\bibinfo{year}{1988}).

\bibitem[{\citenamefont{Mitchell et~al.}(2017)\citenamefont{Mitchell, Koning,
  Vitelli, and Irvine}}]{mitchell2016}
\bibinfo{author}{\bibfnamefont{N.~P.} \bibnamefont{Mitchell}},
  \bibinfo{author}{\bibfnamefont{V.}~\bibnamefont{Koning}},
  \bibinfo{author}{\bibfnamefont{V.}~\bibnamefont{Vitelli}}, \bibnamefont{and}
  \bibinfo{author}{\bibfnamefont{W.}~\bibnamefont{Irvine}},
  \bibinfo{journal}{Nat. Mater.} \textbf{\bibinfo{volume}{16}},
  \bibinfo{pages}{89} (\bibinfo{year}{2017}).

\bibitem[{\citenamefont{Yao}(2017{\natexlab{b}})}]{yao2017emergent}
\bibinfo{author}{\bibfnamefont{Z.}~\bibnamefont{Yao}}, \bibinfo{journal}{Phys.
  Rev. E} \textbf{\bibinfo{volume}{96}}, \bibinfo{pages}{062139}
  (\bibinfo{year}{2017}{\natexlab{b}}).

\bibitem[{\citenamefont{Hwang et~al.}(2019)\citenamefont{Hwang, Weitz, and
  Spaepen}}]{hwang2019direct}
\bibinfo{author}{\bibfnamefont{H.}~\bibnamefont{Hwang}},
  \bibinfo{author}{\bibfnamefont{D.~A.} \bibnamefont{Weitz}}, \bibnamefont{and}
  \bibinfo{author}{\bibfnamefont{F.}~\bibnamefont{Spaepen}},
  \bibinfo{journal}{Proc. Natl. Acad. Sci. U.S.A.}
  \textbf{\bibinfo{volume}{116}}, \bibinfo{pages}{1180} (\bibinfo{year}{2019}).

\bibitem[{\citenamefont{Vicsek and Zafeiris}(2012)}]{vicsek2012collective}
\bibinfo{author}{\bibfnamefont{T.}~\bibnamefont{Vicsek}} \bibnamefont{and}
  \bibinfo{author}{\bibfnamefont{A.}~\bibnamefont{Zafeiris}},
  \bibinfo{journal}{Phys. Rep.} \textbf{\bibinfo{volume}{517}},
  \bibinfo{pages}{71} (\bibinfo{year}{2012}).

\bibitem[{\citenamefont{Schaller and Bausch}(2013)}]{schaller2013topological}
\bibinfo{author}{\bibfnamefont{V.}~\bibnamefont{Schaller}} \bibnamefont{and}
  \bibinfo{author}{\bibfnamefont{A.~R.} \bibnamefont{Bausch}},
  \bibinfo{journal}{Proc. Natl. Acad. Sci. U.S.A.}
  \textbf{\bibinfo{volume}{110}}, \bibinfo{pages}{4488} (\bibinfo{year}{2013}).

\bibitem[{\citenamefont{Nguyen et~al.}(2014)\citenamefont{Nguyen, Klotsa,
  Engel, and Glotzer}}]{nguyen2014emergent}
\bibinfo{author}{\bibfnamefont{N.~H.} \bibnamefont{Nguyen}},
  \bibinfo{author}{\bibfnamefont{D.}~\bibnamefont{Klotsa}},
  \bibinfo{author}{\bibfnamefont{M.}~\bibnamefont{Engel}}, \bibnamefont{and}
  \bibinfo{author}{\bibfnamefont{S.~C.} \bibnamefont{Glotzer}},
  \bibinfo{journal}{Phys. Rev. Lett.} \textbf{\bibinfo{volume}{112}},
  \bibinfo{pages}{075701} (\bibinfo{year}{2014}).

\bibitem[{\citenamefont{Yao}(2019)}]{yao2019command}
\bibinfo{author}{\bibfnamefont{Z.}~\bibnamefont{Yao}}, \bibinfo{journal}{Phys.
  Rev. Lett.} \textbf{\bibinfo{volume}{122}}, \bibinfo{pages}{228002}
  (\bibinfo{year}{2019}).

\bibitem[{\citenamefont{Fermi et~al.}(1955)\citenamefont{Fermi, Pasta, Ulam,
  and Tsingou}}]{fermi1955studies}
\bibinfo{author}{\bibfnamefont{E.}~\bibnamefont{Fermi}},
  \bibinfo{author}{\bibfnamefont{P.}~\bibnamefont{Pasta}},
  \bibinfo{author}{\bibfnamefont{S.}~\bibnamefont{Ulam}}, \bibnamefont{and}
  \bibinfo{author}{\bibfnamefont{M.}~\bibnamefont{Tsingou}},
  \bibinfo{type}{Tech. Rep.}, \bibinfo{institution}{Los Alamos Scientific Lab.}
  (\bibinfo{year}{1955}).

\bibitem[{\citenamefont{Saporta~Katz and
  Efrati}(2019)}]{PhysRevLett.122.024102}
\bibinfo{author}{\bibfnamefont{O.}~\bibnamefont{Saporta~Katz}}
  \bibnamefont{and} \bibinfo{author}{\bibfnamefont{E.}~\bibnamefont{Efrati}},
  \bibinfo{journal}{Phys. Rev. Lett.} \textbf{\bibinfo{volume}{122}},
  \bibinfo{pages}{024102} (\bibinfo{year}{2019}).

\bibitem[{\citenamefont{Saporta~Katz and Efrati}(2020)}]{PhysRevE.101.032211}
\bibinfo{author}{\bibfnamefont{O.}~\bibnamefont{Saporta~Katz}}
  \bibnamefont{and} \bibinfo{author}{\bibfnamefont{E.}~\bibnamefont{Efrati}},
  \bibinfo{journal}{Phys. Rev. E} \textbf{\bibinfo{volume}{101}},
  \bibinfo{pages}{032211} (\bibinfo{year}{2020}).

\bibitem[{\citenamefont{Maxwell}(1860)}]{maxwell1860v}
\bibinfo{author}{\bibfnamefont{J.~C.} \bibnamefont{Maxwell}},
  \bibinfo{journal}{Philos. Mag.} \textbf{\bibinfo{volume}{19}},
  \bibinfo{pages}{19} (\bibinfo{year}{1860}).

\bibitem[{sup()}]{supp2022}
\bibinfo{howpublished}{See Supplemental Material for technical details of the
  numerical approach, and supplemental information about trajectory evolution,
  frequency spectrum analysis of kinetic energy and statistical analysis of
  particle trajectories.}

\bibitem[{\citenamefont{Rapaport}(2004)}]{rapaport2004art}
\bibinfo{author}{\bibfnamefont{D.}~\bibnamefont{Rapaport}},
  \emph{\bibinfo{title}{The Art of Molecular Dynamics Simulation}}
  (\bibinfo{publisher}{Cambridge University Press, Cambridge, UK},
  \bibinfo{year}{2004}).

\bibitem[{\citenamefont{Yao}(2021)}]{yao2021epl}
\bibinfo{author}{\bibfnamefont{Z.}~\bibnamefont{Yao}},
  \bibinfo{journal}{Europhys. Lett.} \textbf{\bibinfo{volume}{133}},
  \bibinfo{pages}{54002} (\bibinfo{year}{2021}).

\bibitem[{\citenamefont{Sundararajan}(2001)}]{sundararajan2001discrete}
\bibinfo{author}{\bibfnamefont{D.}~\bibnamefont{Sundararajan}},
  \emph{\bibinfo{title}{The Discrete Fourier Transform: Theory, Algorithms and
  Applications}} (\bibinfo{publisher}{World Scientific}, \bibinfo{year}{2001}).

\bibitem[{\citenamefont{Altmann et~al.}(2006)\citenamefont{Altmann, Motter, and
  Kantz}}]{altmann2006stickiness}
\bibinfo{author}{\bibfnamefont{E.~G.} \bibnamefont{Altmann}},
  \bibinfo{author}{\bibfnamefont{A.~E.} \bibnamefont{Motter}},
  \bibnamefont{and} \bibinfo{author}{\bibfnamefont{H.}~\bibnamefont{Kantz}},
  \bibinfo{journal}{Phys. Rev. E} \textbf{\bibinfo{volume}{73}},
  \bibinfo{pages}{026207} (\bibinfo{year}{2006}).

\bibitem[{\citenamefont{Chirkin et~al.}(2004)\citenamefont{Chirkin, Novikov,
  and Laptev}}]{chirkin2004nonclassical}
\bibinfo{author}{\bibfnamefont{A.~S.} \bibnamefont{Chirkin}},
  \bibinfo{author}{\bibfnamefont{A.~A.} \bibnamefont{Novikov}},
  \bibnamefont{and} \bibinfo{author}{\bibfnamefont{G.~D.}
  \bibnamefont{Laptev}}, \bibinfo{journal}{J. Opt. B}
  \textbf{\bibinfo{volume}{6}}, \bibinfo{pages}{S483} (\bibinfo{year}{2004}).

\bibitem[{\citenamefont{Landau and Lifshitz}(1976)}]{Landau_mechanics}
\bibinfo{author}{\bibfnamefont{L.~D.} \bibnamefont{Landau}} \bibnamefont{and}
  \bibinfo{author}{\bibfnamefont{E.}~\bibnamefont{Lifshitz}},
  \emph{\bibinfo{title}{Mechanics, 3rd Edition}}
  (\bibinfo{publisher}{Butterworth-Heinemann,Oxford}, \bibinfo{year}{1976}).

\bibitem[{\citenamefont{Feigenbaum}(1978)}]{Feigenbaum1978}
\bibinfo{author}{\bibfnamefont{M.}~\bibnamefont{Feigenbaum}},
  \bibinfo{journal}{J. Stat. Phys.} \textbf{\bibinfo{volume}{19(1)}},
  \bibinfo{pages}{25} (\bibinfo{year}{1978}).

\bibitem[{\citenamefont{De~Wijn and Fasolino}(2009)}]{de2009relating}
\bibinfo{author}{\bibfnamefont{A.~S.} \bibnamefont{De~Wijn}} \bibnamefont{and}
  \bibinfo{author}{\bibfnamefont{A.}~\bibnamefont{Fasolino}},
  \bibinfo{journal}{J. Phys. Condens. Matter} \textbf{\bibinfo{volume}{21}},
  \bibinfo{pages}{264002} (\bibinfo{year}{2009}).

\bibitem[{\citenamefont{Saari and Xia}(1995)}]{saari1995off}
\bibinfo{author}{\bibfnamefont{D.~G.} \bibnamefont{Saari}} \bibnamefont{and}
  \bibinfo{author}{\bibfnamefont{Z.}~\bibnamefont{Xia}},
  \bibinfo{journal}{Notices of the AMS} \textbf{\bibinfo{volume}{42}},
  \bibinfo{pages}{538–546} (\bibinfo{year}{1995}).

\bibitem[{\citenamefont{Rom-Kedar and Turaev}(2022)}]{rom2022stable}
\bibinfo{author}{\bibfnamefont{V.}~\bibnamefont{Rom-Kedar}} \bibnamefont{and}
  \bibinfo{author}{\bibfnamefont{D.}~\bibnamefont{Turaev}},
  \bibinfo{journal}{arXiv:2208.14993}  (\bibinfo{year}{2022}).

\bibitem[{\citenamefont{Gaveau and Schulman}(2015)}]{epjst2015}
\bibinfo{author}{\bibfnamefont{B.}~\bibnamefont{Gaveau}} \bibnamefont{and}
  \bibinfo{author}{\bibfnamefont{L.~S.} \bibnamefont{Schulman}},
  \bibinfo{journal}{Eur. Phys. J. Spec. Top.} \textbf{\bibinfo{volume}{224}},
  \bibinfo{pages}{891} (\bibinfo{year}{2015}).

\end{thebibliography}

\end{document}